\pdfoutput=1
\documentclass[epj]{pw}

\usepackage{amssymb}
\usepackage{graphics}
\usepackage{graphicx}
\usepackage{amsmath}

\usepackage[usenames,dvipsnames]{color}

\begin{document}
\title{  Inhomogeneous Light Shift Effects on Atomic Quantum State Evolution in Non-Destructive Measurements}
\author{Patrick Windpassinger\thanks{\email{pwindpas@nbi.dk}}   \and Daniel Oblak   \and Ulrich Busk Hoff \and J\"{u}rgen Appel \and Niels Kj{\ae}rgaard \and Eugene S. Polzik
}
\authorrunning{P. Windpassinger {\it et al.} }
\titlerunning{Inhomogeneous Light Shift Effects on Quantum State
Evolution}
\institute{QUANTOP, Niels Bohr Institute, University of Copenhagen,
Denmark}
\date{Dated: January 18, 2008 }
\abstract{Various parameters of a trapped collection of cold and
ultracold atoms can be determined non--destructively by measuring
the phase shift of an off--resonant probe beam, caused by the  state
dependent index of refraction of the atoms.  The dispersive
light--atom interaction, however, gives rise to a differential light
shift (AC Stark shift) between the atomic states which, for a
nonuniform probe intensity distribution, causes an inhomogeneous
dephasing between the atoms. In this paper, we investigate the
effects of this inhomogeneous light shift in non--destructive
measurement schemes. We interpret our experimental data on
dispersively probed Rabi oscillations and Ramsey fringes in terms of
a simple light shift model which is shown to describe the observed
behavior well. Furthermore, we show that by using spin echo
techniques, the inhomogeneous phase shift distribution between the
two clock levels can be reversed.
\PACS{
      {32.80.-t}{photon--atom interactions}\and
      {03.65.Yz}{decoherence, quantum mechanics}   \and
      {06.30.Ft}{clocks}
     }
}
\maketitle
\section{Introduction}\label{intro}
Resonant absorption and fluorescence measurements have been employed
extensively in recent years to probe the properties of cold and
ultracold atomic gasses. For example, Bose Einstein condensates are
typically recorded in absorption imaging. Resonant light--atom
interaction, however, destroys initial sample properties such as
coherences between the internal states of the atoms. Phase contrast
imaging using off resonant light offers an alternative,
non-destructive means of probing, which has proven  viable, e.g. in
following the evolution of the vector magnetization density with
repeated imaging of the same atomic sample \cite{Sadler2006}. In
general, such measurements can be used to probe the evolution of the
sample density   \cite{Petrov2007} and the internal state
populations \cite{Chaudhury2006,Usami2007,Windpassinger2007}, When
fulfilling the requirements for  Quantum Nondemolition (QND)
measurements \cite{Kuzmich1998}, the dispersive interaction,
furthermore, provides a vehicle for quantum state generation in
ensembles of atoms, for example for the generation of spin squeezed
and entangled states of atoms \cite{Julsgaard2001,Geremia2004}. Such
QND measurements form a basis for a number of protocols in the field
of quantum information science \cite{Sherson2006,Hammerer2004} such
as quantum memory \cite{Julsgaard2004}, and quantum teleportation
\cite{ShersonNat2006}.\\
Non-destructive measurements could potentially prove useful in the
operation of atomic clocks. The accuracy of state--of--the art
clocks is presently limited by the projection noise
\cite{Santarelli1999}, which arises from the probabilistic
uncertainty associated with a projective measurement  of independent
particles in a quantum mechanical superposition state. To reduce
this uncertainty, one can explore the possibility of creating so
called squeezed states
\cite{Wineland1992,Wineland1994,Oblak2005,Meiser2007} via QND
measurements, where the particles are no longer independent but
rather non--classically correlated (entangled). Furthermore, in
approaches like optical lattice clocks, a way of improving the
signal-to-noise ratio, which directly enters into the frequency
stability, is to increase the duty cycle of sample preparation
relative to the actual interrogation time
\cite{Takamoto2005,Targat2006,Ludlow2006}. This can be achieved  by
using non--destructive probing schemes where the trapped sample,
once prepared, is reused several times during the lifetime
of the trap. \\
While off--resonant probing may lead  to negligible decoherence due
to spontaneous photon scattering, the dispersive light--atom
interaction inevitable affects the atomic states via the AC--Stark
shift. The influence of this light shift in Caesium fountain clocks,
where it is  induced by a homogeneously distributed off--resonant
light field, has been studied in \cite{Featonby1998}. The assessment
of such effects and the ability to account for them, is of major
importance when employing non--destructive probing schemes in
practical applications.\\
We have constructed a Caesium atomic clock, using a dipole trapped
cold sample. Our eventual and primary goal is to demonstrate
pseudo--spin squeezing of the clock transition via a QND measurement
\cite{Oblak2005}. To that end, we read out the population of the
clock states non--destructively by measuring the phase shift
acquired by probe light due to the off--resonant index of refraction
of the atomic medium. The phase shift of the probe light is measured
with a Mach--Zehnder interferometer, operating close to the standard
quantum limit \cite{Windpassinger2007}. Since our readout beam has a
Gaussian intensity profile, the interrogation of the (inhomogeneous)
sample induces an inhomogeneous light shift across the atomic
sample, which finally is detected with a non--uniform detection
efficiency. In the present paper we analyze these inhomogeneous
effects by  studying  the evolution of clock--state Rabi
oscillations, and we perform Ramsey spectroscopy measurements  to
further characterize the dephasing. We finally investigate the
reversibility of the probe--introduced dephasing with spin echo
techniques.

\section{Experimental setup}
\subsection{Framework}
\begin{figure}[b!]\begin{center}
\includegraphics[width=\columnwidth]{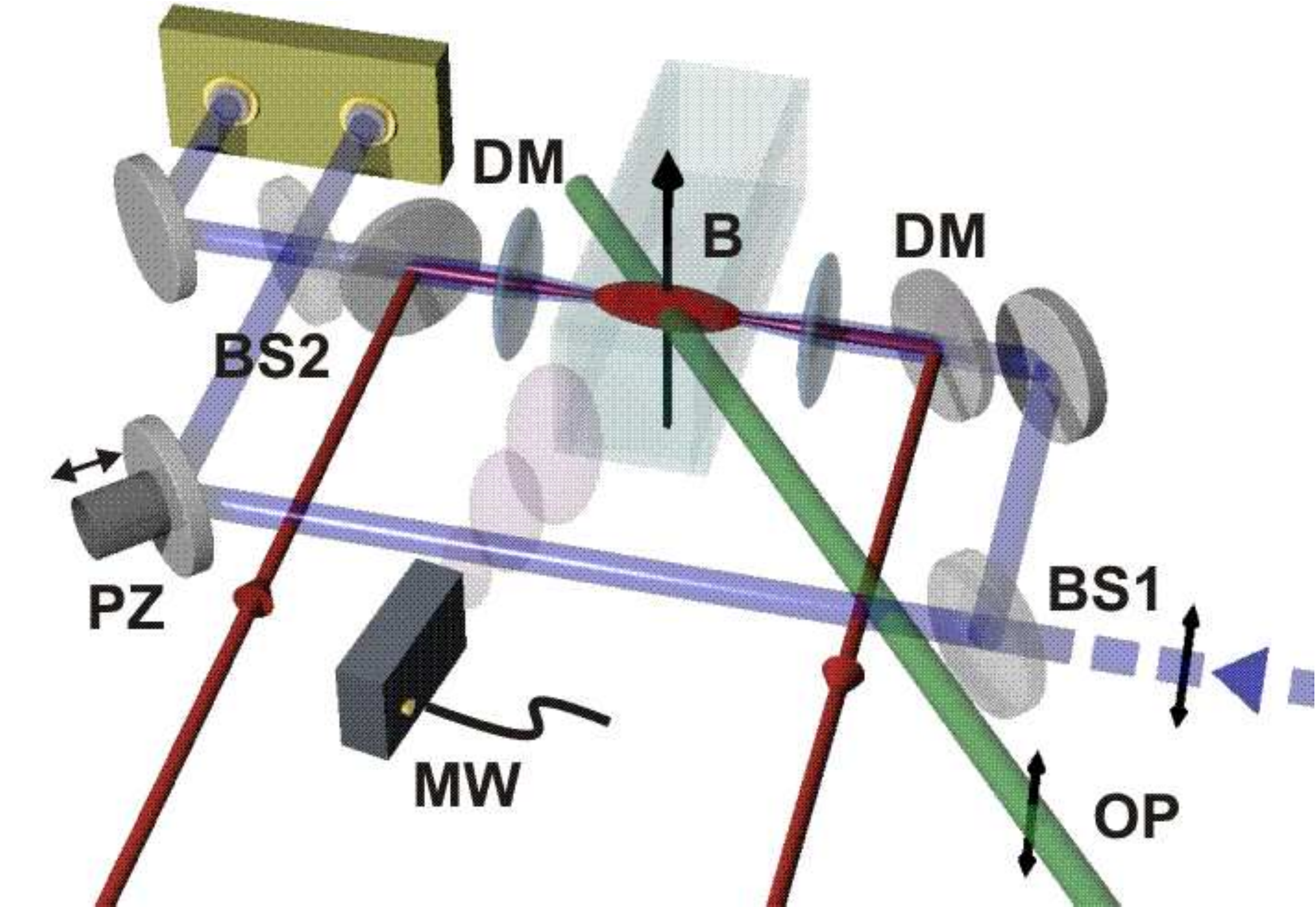}
\caption{Schematic drawing of the experimental setup. A probe laser
beam enters the interferometer via a port of the beam splitter BS1.
One part of the beam propagates through the Cs atoms confined by a
dipole trap, as they interact with the radiation field of the
microwave source MW.  The dipole beam is coupled into the probe arm
via two dichroic mirrors DM. The other part of the beam propagates
on a reference path with no atoms. The beams are recombined at the
beam splitter BS2 and the phase shift from the atoms is detected as
the differential signal of detectors at the two output ports of BS2.
The geometrical path length difference of the interferometer arms is
stabilized with a piezoelectric element PZ. B shows the direction of
the magnetic guiding field and OP the two counter--propagating
optical pumping beams. }\label{setup}\end{center}
\end{figure}

\begin{figure}
\begin{center}
\resizebox{.9\columnwidth}{!}{  \includegraphics{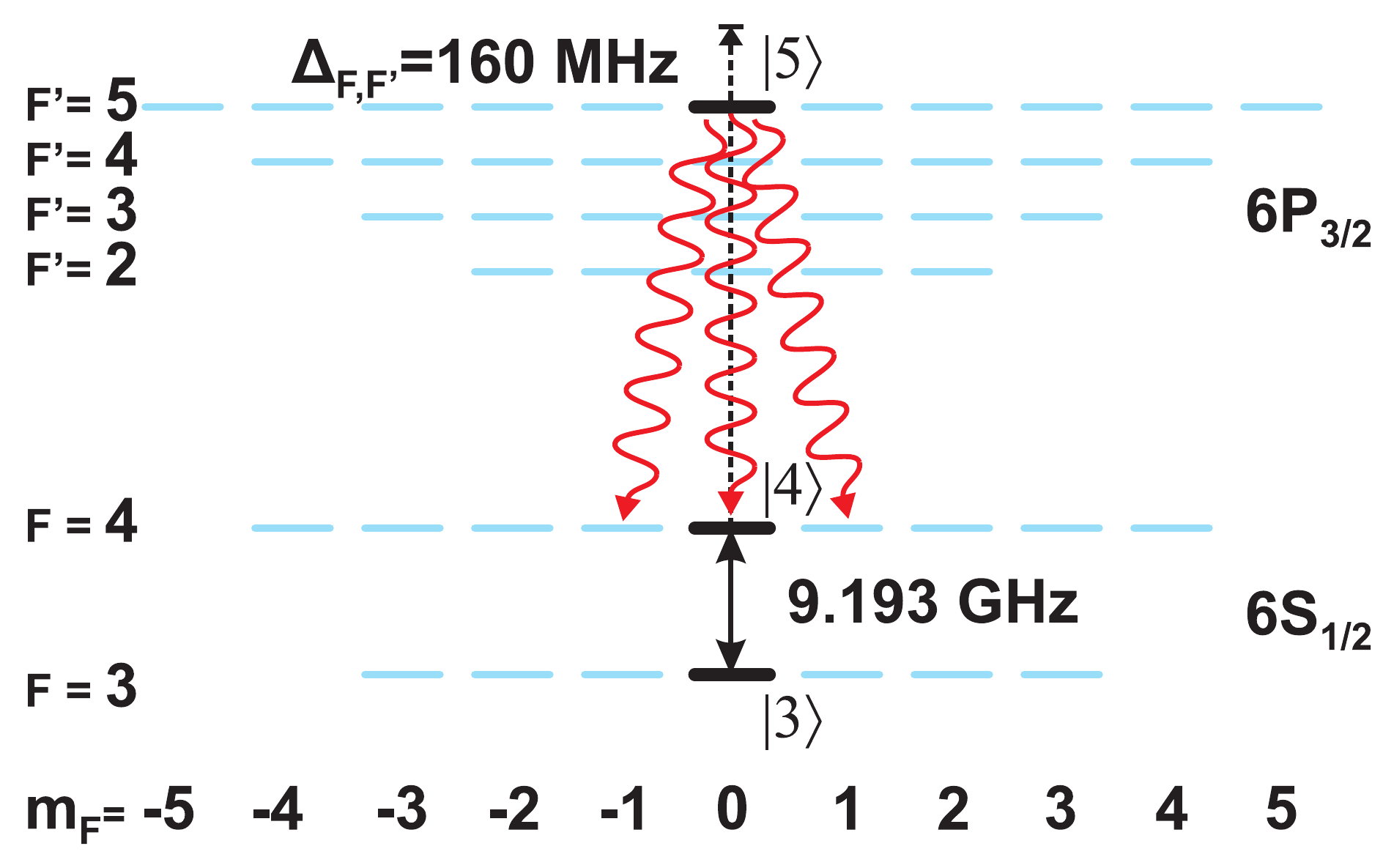} }
\caption{Schematic level diagram indicating a resonant drive between
the states $|3\rangle$ and $|4\rangle$ as well as dispersive
(off-resonant) coupling between the states $|4\rangle$ and
$|5\rangle$. In the experiment $|3\rangle$ is the
$6S_{1/2}(F=3,m_F=0)$ Cs clock state, $|4\rangle$ is the
$6S_{1/2}(F=4,m_F=0)$ clock state, and $|5\rangle$ is the
$6P_{3/2}(F'=5,m_F=0)$ state. We also indicate three possible
spontaneous photon scattering channels.} \label{fig:1}\end{center}
\end{figure}

The system we are considering is the standard microwave clock
transition $6S_{1/2}(F=3,m_F=0)\equiv |3\rangle \leftrightarrow
6S_{1/2}(F=4,m_F=0)\equiv |4\rangle$ \cite{Vanier1989} in cold
Cs--atoms. Earlier versions of the experimental setup have been
described in \cite{Oblak2005,Petrov2007,Windpassinger2007}. A
schematic drawing of the setup is shown in Fig. \ref{setup}.
A typical experimental cycle starts by loading Cs atoms into a
magneto--optical trap and after a sub-Doppler cooling stage, we
transfer about $10^5$ Cs atoms with a temperature of $\sim 15\,\mu$K
into a single beam, far-off resonance dipole trap \cite{Grimm2000}.
A diode pumped Yb:YAG disk laser at 1032\,nm produces 4\,W of dipole
trapping beam, focussed down to a waist of about $50\,\mu$m. To
initialize the atomic ensemble to one of the clock states, a
homogeneous, magnetic guiding field of $\sim 1$\,Gauss is applied
and the atoms are subsequently optically pumped into the
$6S_{1/2}(F=4,m_F=0)$ ground state by simultaneously applying
linearly polarized light to the $6S_{1/2}(F = 4) \to 6P_{3/2}(F'=4)$
and $6S_{1/2}(F = 3) \to 6P_{3/2}(F'=4)$ transitions
\cite{AVILA1987}. A schematic level diagram of Cs is shown in Fig.
\ref{fig:1} for reference.
The efficiency of the optical pumping is limited to about 80\% by
off--resonant excitations, magnetic background field fluctuations,
polarization impurities and the narrow bandwidth of the pumping
laser \cite{Tremblay1990}. To achieve a purer ensemble, where all
atoms populate a single clock state, we first transfer the
population in the $|4\rangle$ state to the $|3\rangle$ clock state
with a resonant microwave $\pi$--pulse. The remaining atoms in
$6S_{1/2}(F=4,m_F\neq 0)$ are then expelled from the trap with the
resonant light on the cycling $6S_{1/2}(F=4) \leftrightarrow
6P_{3/2}(F'=5)$ transition. In this way we obtain an
ensemble with $\gtrsim 99\,\%$ purity. \\
To drive a coherent evolution of the clock state populations, $
|3\rangle \leftrightarrow |4\rangle$, we apply a linearly polarized
microwave field with a frequency of $\sim 9.2$\,GHz with various
durations and powers to the atoms. The microwave field is generated
by a HP8341B precision synthesizer and amplified with a solid state
amplifier to 1W. To stabilize the power, we split a portion of the
microwave power into a solid state detector and feed the signal
directly back onto the synthesizer's external stabilization input.
Pulse shaping is done with a HP4720A pulse modulator inserted after
the feedback loop and the resulting microwave pulses are directed
into the vacuum cell via a cut--to--size rectangular waveguide. The
microwaves produce the Rabi flopping, characteristic to a two level
system strongly driven  with a near resonant coupling field
\cite{Tannoudji1977}.
\\
To read out the population  of the ensemble in the state $|4\rangle$
non--de\-struc\-tively, we measure the phase shift of the probe
light \cite{Oblak2005,Petrov2007} caused by the state dependent
refractive index of atoms \cite{Loudon1973}:
\begin{eqnarray}
  \Delta \phi = \frac{5}{36}\phi_0  N_{|4\rangle}   \frac{
    (\gamma/2)\Delta_{F,F'}}{\Delta_{F,F'}^2 + (\gamma/2)^2},
    \label{phaseshiftformulaspec}
\end{eqnarray}
where $\phi_0=3l\lambda^2/\pi V$,  $N_{|4\rangle} $ is the number of
atoms in the ${|4\rangle} $ clock state, $\Delta_{F,F'}$ is the
detuning of the probe light from the $6S_{1/2}(F = 4) \to
6P_{3/2}(F' = 5)$ transition, $\lambda$ is the wavelength of the
probe light, $\gamma$ is the linewidth of the transition, and $l$ is
the length and $V$ the volume of the sample. To obtain the reduced
form of the phase shift, given in equation
(\ref{phaseshiftformulaspec}), we have assumed a pure sample in
${|4\rangle} $ and  neglected
couplings to the $6P_{3/2}(F' \neq5$) levels \cite{Windpassinger2007}.  \\
Experimentally, the phase shift of the probe light is recorded by
placing the ensemble into one arm of a Mach--Zehnder interferometer
and detecting the two interferometer outputs with two photodiodes
whose outputs are fed into a low noise AC integrating
photoamplifier. The differential output of the detector is directly
digitized with a storage oscilloscope, saved to disk and the
recorded pulses are  numerically integrated afterwards. For the
probe light, we arrange the detuning  $\Delta_{F,F'} = + 160\,$MHz
such that we get a considerable phase shift of up to half a radian
due to the presence of atoms in the interferometer, while keeping
the probability of spontaneous photon scattering low. With $\sim
10^5$ photons in one probe pulse, the spontaneous scattering
probability per atom is around $0.05$\,\%. The pulses are generated
using a standard acousto--optical modulator and with typical
durations between 200\,ns and a few microseconds. To clean the
transverse mode of the beam before entering the interferometer, the
pulsed probe--beam is coupled into an optical fibre. The path length
difference of the two interferometer arms is actively stabilized
against thermal drifts and acoustic noise by applying feedback to
one of the folding mirrors. The error signal for the feedback is
obtained by matching a weak, pulsed locking laser with $\sim
1.5\,\mu$W equivalent DC power into the mode cleaning fibre used for
the probe pulses, and demodulating the signal from the
photo--receiver. We usually lock the interferometer to the white
light position, where both arms have the same optical path length.
To eliminate an influence of the locking laser onto the atoms, its
wavelength is $\sim 20$\,nm blue detuned from the $6S_{1/2} \to
6P_{3/2}$ transition. Due to its detuning, the locking beam is not
affected by the presence of atoms so that the {\it geometrical} path
length can be fixed irrespectively of the atomic state and density.\\

Using our non--destructive probing scheme, we can follow the
evolution of the atomic ensemble quantum state when subjected to
external fields. More specifically, we measure the population in the
$|4\rangle$ clock state.  Figure \ref{figrabi}a shows a typical
recording of microwave induced Rabi oscillations on the clock
transition. A constant resonant microwave driving  field is applied
while the atomic ensemble is probed every $10.3\,\mu$s with
$0.2\,\mu$s optical probe pulses, corresponding to $\sim 10^5$
photons per pulse. The figure represents an average of 10
experimental runs each sampling the atoms $\sim 500$ times
\cite{Windpassinger2007}. From a fit to the data, assuming a cosine
function with  exponentially decaying envelope, we extract a time
constant of $\tau=3.0$\,ms.
\subsection{Motivation}
\begin{figure*}[t!]\begin{center}
\includegraphics[width=2\columnwidth]{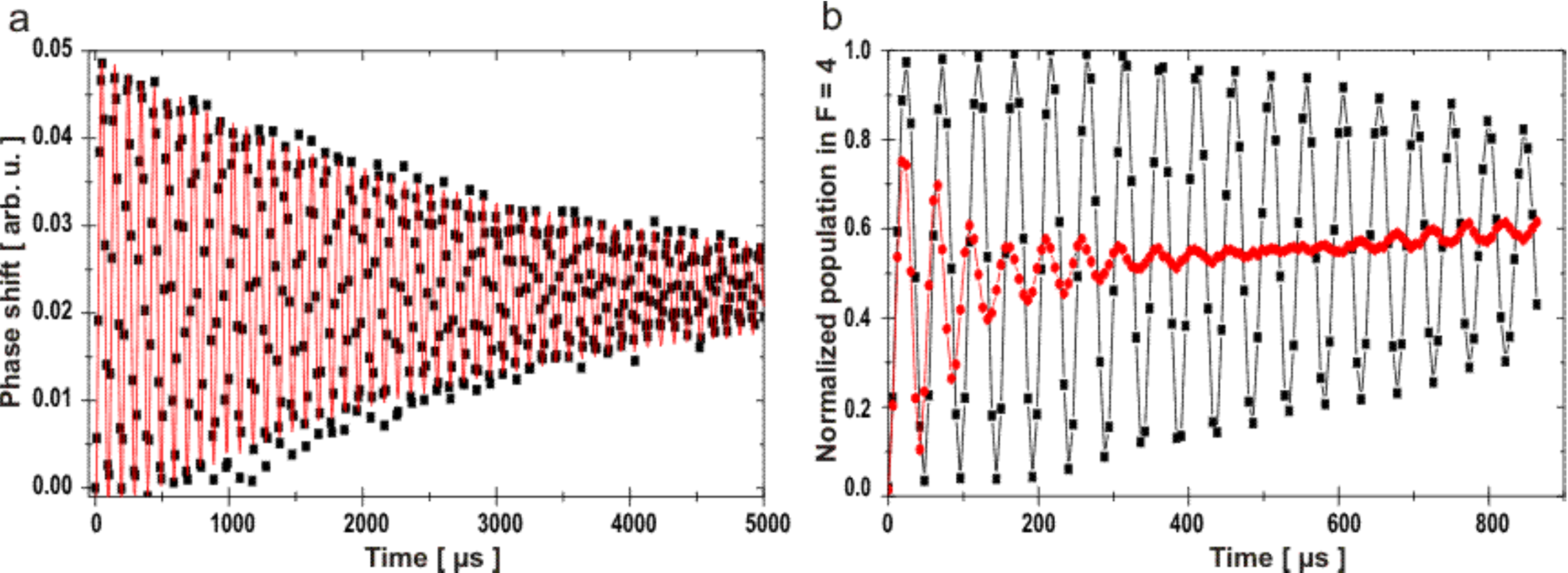}\end{center}
\caption{(a) Non--destructively probed Rabi oscillations. Each probe
pulse contains about $10^5$ photons, is $0.2\,\mu$s long and the
repetition period of the pulses is $10.3\,\mu$s. The data
corresponds to an average of 10 runs of the experiment, each
sampling the atoms  $\sim 500$ times.  The solid line corresponds to
a fit, using a cosine function with exponential amplitude damping.
(b) Changing the probe strength from $1.5 \times 10^5$ photons per
pulse (data indicated with {\tiny $\blacksquare$}) by a moderate
factor of 4, a drastic change in the Rabi oscillation's envelope
occurs (data indicated by {\color{red}$ \bullet$}). The solid lines
are to guide the eye, only. } \label{figrabi}
\end{figure*}

Some care is required when describing dispersive probing as
non--destructive. In our case, the probing is non--destructive in
the sense that the spontaneous photon scattering is kept very low.
With the exception of the case where the ensemble is in one of the
eigenstates $|3\rangle$ or $|4\rangle$, spontaneous scattering
events destroy coherences irreversibly \cite{Ozeri2005}, i.e.
project superposition states onto eigenstates. However, even for
negligible spontaneous scattering, the atomic quantum state will be
affected: The dispersive light--atom interaction will introduce a
phase shift between the atomic states $|3\rangle$ and $|4\rangle$
due to the light shift caused by the probe \cite{Featonby1998}. When
probing Rabi oscillations non-destructively, we observe a very
distinct change in the envelope when changing the probe power rather
moderately. Figure \ref{figrabi}b shows two traces of Rabi
oscillations recorded with the same probe pulse duration of
$1.0\,\mu$s and the $7\,\mu$s repetition period but with different
probe photon numbers.  With $1.5 \times 10^5$ photons per pulse we
obtain a decay constant of $\tau = 1.5$\,ms, which reduces to $\tau
= 80\,\mu$s when the photon number per pulse is increased to $6.3
\times 10^5$. This much faster decay cannot be explained just by the
four times higher spontaneous excitation probability. Additionally,
the trace corresponding to the higher probe power shows a clear
revival of the oscillations at around $t = 800\,\mu$s, and a change
in the Rabi oscillation frequency is observable. In the following
sections, we systematically analyze these effects and demonstrate
that they can be well understood when taking the spatial
inhomogeneity of the probe beam into account which causes a spatial
distribution of the differential light shift between the clock
states.

\section{Inhomogeneous dephasing due to probe light shift} \label{sec:2}
\subsection{Bloch sphere picture and interpretation}\label{sec:2:1}
We model the column density of the atomic sample on polar
coordinates $(r,\phi)$ by $n(r)=n_0 \exp(-2r^2/r_0^2)$, where $n_0$
is the peak column density and $r_0$ characterizes the sample
radius. The atomic sample interacts with a Gaussian laser beam
propagating along the $z$-axis and focussed at the sample's
location. In the case when the axial size of the atomic sample is
short compared to the Rayleigh range of the laser beam we can
approximate the light intensity distribution within the interaction
volume by $I(r)=I_0 \exp(-2r^2/w_0^2)$, where $I_0$ is the peak
intensity and $w_0$ is the Gaussian beam waist.

The atoms in the sample are described by the two internal states
$|3\rangle$ and $|4\rangle$ separated by an energy $E_{34}=\hbar
\omega_{34}$ and on-resonance Rabi oscillations between the states
can be induced by a drive field of angular frequency $\omega_{34}$
\cite{Tannoudji1977}. The laser beam is assumed to have a strong
dispersive interaction with atoms in the $|4\rangle$ and negligible
coupling to the $|3\rangle$ state. Hence, the population in the
$|4\rangle$ state can be recorded as a phase shift of the laser
beam. In turn, the dispersive interaction adds a differential light
shift  $\Delta E$ to $E_{34}$ (e.g.\cite{Metcalf1999}).

The dynamics as a result of  the combined action of continuously
driving the $|3\rangle\leftrightarrow |4\rangle$ transition and
acting on the sample with pulses of laser light at given instances
of time is conveniently described in the Bloch sphere picture. A
general superposition state $\cos\frac{\theta}{2}|3\rangle + e^{i
\phi} \sin\frac{\theta}{2}|4\rangle$ is, up to a global phase,
mapped onto the unit sphere at polar angle $\theta$ and azimuthal
angle $\phi$ \cite{Allen1987}. An ensemble state {$\bf s$} is thus
fully described by the two angles $ \theta $ and $ \phi$ or by its
cartesian components ${\bf s} = s_1 {\bf e}_1+ s_2 {\bf e}_2 +s_3
{\bf e}_3$. Pure states with all the atoms in $|3\rangle$ or
$|4\rangle$ correspond to the Bloch vectors ${\bf s}=(0,0,-1)$ or
${\bf s}=(0,0,1)$, respectively. Under the influence of a resonant
driving field of duration $t$, the initial atomic state {$\bf s_0$}
transforms according to
\begin{equation}{\bf s}(t) =  \bar{\bar{U}}(t)
{\bf s}_0
\end{equation}
\begin{equation}
\label{mwrot} \bar{\bar{ U}}_{\mathrm {drive}}(t) =   \left[
            \begin{array}{ccc}
              1 & 0 & 0 \\
              0 & \cos\Omega t & \sin \Omega t\\
              0 & -\sin \Omega t& \cos \Omega t\\
            \end{array}
          \right],
\end{equation}
where $\Omega$ is the  resonant Rabi frequency. Similarly, the
dispersive interaction for a time $t$ is described by the evolution
matrix
\begin{equation}\label{mwprobe}
    \bar{\bar{U}}_{\mathrm {probe}}(t)=\left[
            \begin{array}{ccc}
              \cos\chi t & \sin \chi t & 0 \\
              -\sin \chi t& \cos \chi t  & 0\\
              0 & 0 & 1 \\
            \end{array}
          \right],
\end{equation}
where $\chi=\Delta E/\hbar$. Since the problem is symmetric, the
same rotation is induced by detuning the driving field from
resonance by $\Delta E$ for a time $t$ or changing the phase of the
driving field by $\chi t$. Finally, for the simultaneous action of
the two effects we obtain the evolution matrix $
\bar{\bar{U}}_{\mathrm { combined}}(t)$:
\begin{equation}\label{mwboth}
   \left[
            \begin{array}{ccc}
              \frac{\Omega^2+\chi^2\cos\sqrt{\Omega^2+\chi^2} t}{{\Omega^2+\chi^2}} & \frac{\chi\sin \sqrt{\Omega^2+\chi^2} t}{\sqrt{\Omega^2+\chi^2}} & \frac{\Omega\chi(1-\cos \sqrt{\Omega^2+\chi^2} t)}{\Omega^2+\chi^2} \\
              \frac{-\chi\sin \sqrt{\Omega^2+\chi^2} t}{\sqrt{\Omega^2+\chi^2}}& \cos \sqrt{\Omega^2+\chi^2} t  & \frac{\Omega\sin \sqrt{\Omega^2+\chi^2} t}{\sqrt{\Omega^2+\chi^2}}\\
              \frac{\Omega\chi(1-\cos \sqrt{\Omega^2+\chi^2} t)}{\Omega^2+\chi^2} & \frac{-\Omega\sin \sqrt{\Omega^2+\chi^2} t}{\sqrt{\Omega^2+\chi^2}} & \frac{\chi^2+\Omega^2\cos\sqrt{\Omega^2+\chi^2} t}{{\Omega^2+\chi^2}} \\
            \end{array}
          \right]
\end{equation}
The evolution of an initial Bloch vector $\bf s_0$ can now be
propagated by multiplying the corresponding matrices, e.g.
(\ref{mwrot}) and (\ref{mwboth}) in succession to get an overall
transfer matrix $T(\chi,t)$. In the experiment, we only have direct
access to the $3$--projection of the Bloch vector.

Due to the Gaussian intensity profile of the laser beam, we get a
position dependent light shift and $\chi$ will vary radially as
$\propto I(r)$. Moreover, in the detection of the $3$-projection of
$\bf s$, the Gaussian intensity dependence of the laser beam in
conjunction with the Gaussian column density of the atomic sample
gives rise to a signal
\begin{eqnarray}\label{phase shift}
    S_\phi(t)&\propto&\int_0^\infty [T(\chi(r),t){\bf s}]_3 n(r)I(r)r dr \nonumber \\
    &\propto&\int_0^\infty [T(\chi(r),t){\bf s}]_3 n_0 I_0
    e^{-2(\frac{r}{r_0})^2}e^{-2(\frac{r}{w_0})^2}r dr\nonumber \\
    &\propto&\int_0^{\chi_0}[T(\chi,t){\bf s}]_3
    \chi^{(\frac{w_0}{r_0})^2}d\chi,
\end{eqnarray}
where $\chi_0=\chi(r=0)$ corresponds to the maximum light shift at
the center of the Gaussian beam. Hence, for a given ratio between
$r_0$ and $w_0$ the net measured Bloch vector results from
infinitesimal contributions from atoms with light shifts $\chi=[0
...\chi_0]$ carrying a weight $\chi^{(\frac{w_0}{r_0})^2}$. In Fig.
\ref{blochsteps}c we plot the weighting factor for a few values of
the ratio $k= w_0/r_0$. As would be expected, the atoms contributing
to the net Bloch vector have undergone practically the same light
shift close to the maximum $\chi_0$ if the laser beam waist is much
larger than the atomic sample radius $w_0\gg r_0$. At the other
extreme $w_0\ll r_0$, our detected signal will have a uniform
contribution from light shifts in the interval $[0 ...\chi_0]$.

\subsection{Application to Rabi oscillations}
\subsubsection{Expected behavior from the theoretical model}
\begin{figure*}[tb!]
\begin{center}
\includegraphics[width=2\columnwidth]{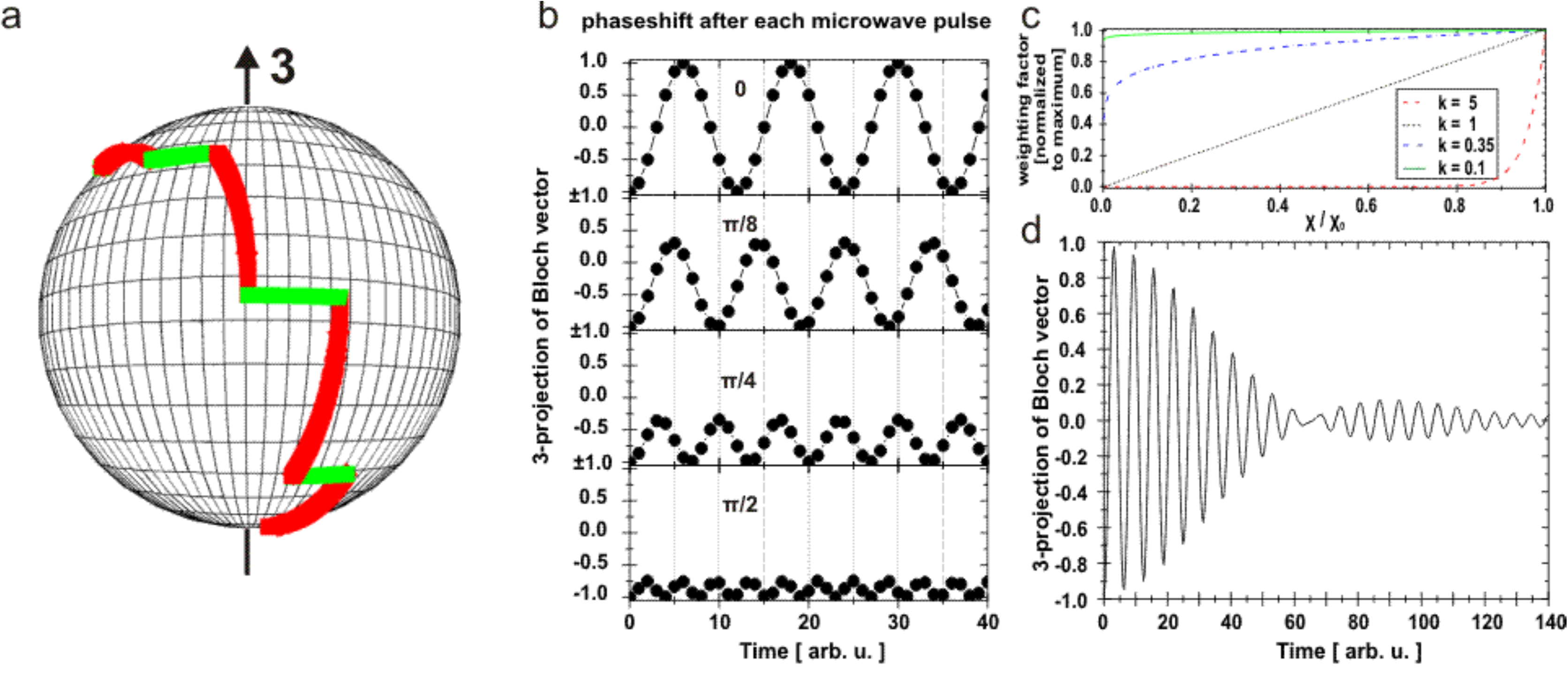}
\end{center}
\caption{(a) State evolution on the Bloch sphere with alternating
microwave pulses, each pulse shifting  by  $\pi/4$ around the
$1$--axis, and probe pulses, each causing a homogeneous shift of
$\pi/9$ around the $3$--axis. (b) 3-- projection of the Bloch vector
for different homogeneous light shifts, applied discretely in
between separated $\pi/6$ microwave pulses. (c) Weighting factor of
the oscillations with different frequency due to the inhomogeneity
of the probe beam and the sample. The factor is plotted for
different ratios $k$ of the probe beam waist $\omega_0$ to the
sample size $r_0$, and normalized to its maximum value. For large
ratios $k \gg 1$, an almost homogeneous shift is induced, for small
ratios $k \ll 1$ the distribution is flat. (d) Rabi--oscillations
resulting from inhomogeneous light shift distribution across the
probe area.} \label{blochsteps}
\end{figure*}

Let us first consider  the theoretical model for the case of
alternating microwave driving field and probe pulses. If we neglect
the inhomogeneity of the induced light shift, each single probe
pulse will cause the tip of the Bloch vector to rotate around the
$3$--axis according to the transformation matrix (\ref{mwprobe}) by
an angle $\chi t$, proportional to the number of photons of the
probe pulse. Alternating microwave pulses, rotating around the
$1$--axis according to matrix (\ref{mwrot}), and probe pulses, we
expect a step--like evolution as shown in Fig. \ref{blochsteps}a.
In figure \ref{blochsteps}b, we show the expected measurement result
for each probe pulse when changing the photon number or the rotation
angle $\chi t$ induced per probe pulse. As can be seen, the
discretely induced transition frequency change $\Delta E = \hbar
\chi$ at discrete times, resulting from the differential light shift
between the clock states, leads  to a higher effective Rabi
frequency $\Omega' = \sqrt{\Omega^2 + \overline{\Delta E} ^2}$. Here
$\overline{\Delta E}=\frac{1}{2 \pi} \int \Delta E(t) dt$ is the
time averaged frequency change. This effect is very similar to the
Rabi frequency change one observes when the transition frequency is
continuously shifted relative to the driving field by $\Delta E$
e.g. due to off--resonant driving \cite{Allen1987} or a homogeneous
light shift across the sample \cite{Chaudhury2006}. Introducing a
light--shift at discrete intervals changes the observed Rabi
frequency stepwise during the single period, however, after each
period  $T= \frac{2 \pi}{\Omega}$ the effect is the same as if the
transition frequency had been changed by a mean value
$\overline{\Delta E}$ during the whole period. In the experiment, a
continuous distribution of light shifts $\chi = [0 \dots \chi_0]$ is
present and thus oscillations of different frequencies, weighted in
amplitude with the density distribution of the sample across the
probe beam, interfere. The resulting oscillations are shown in  Fig.
\ref{blochsteps}d for a probe size to sample ratio $k = 0.35$ and a
maximum shift of $\chi_0=0.3$\,rad per pulse.
\\
\subsubsection{Experimental results}
\begin{figure*}[ht!]
\begin{center}
\includegraphics[width=2\columnwidth]{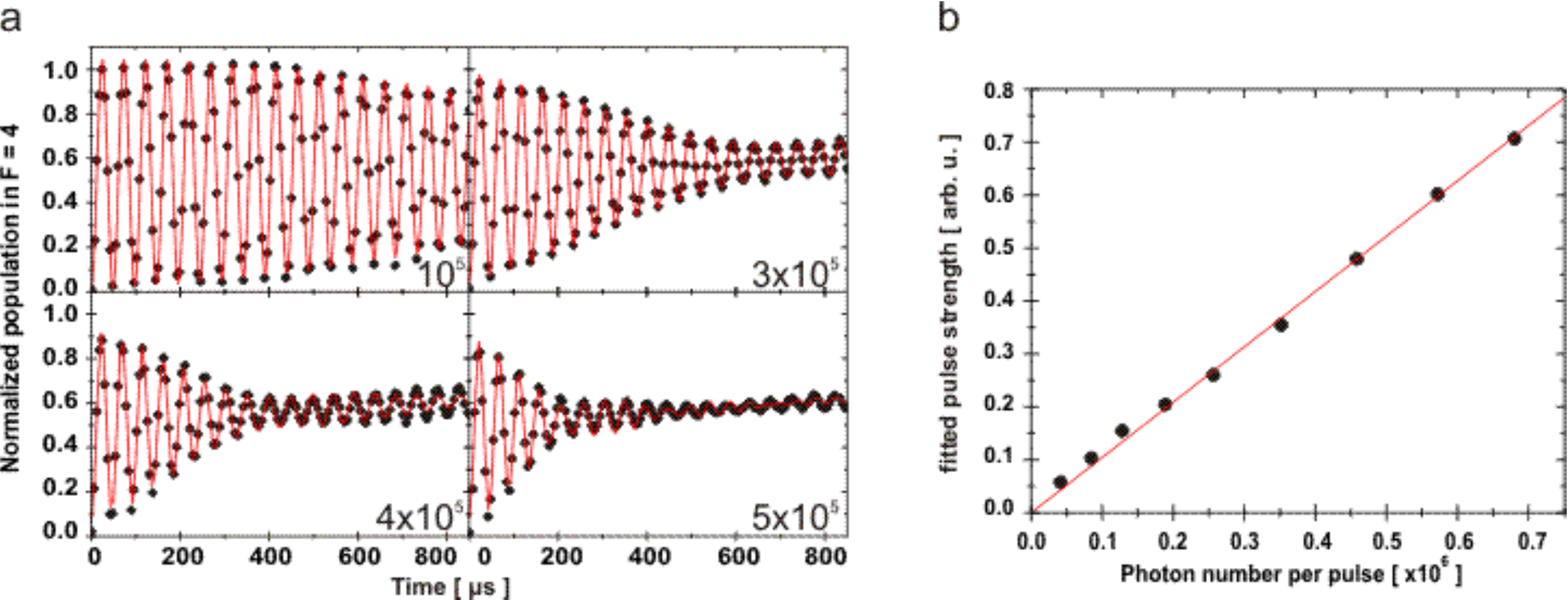}
\end{center}
\caption{(a) Non--destructively probed Rabi oscillations. The photon
number per probe pulse is given in the lower right corner of each
graph. The solid line represents a fit to the data with the model
discussed in the text. (b) Comparison of the photon number per pulse
as measured in the experiment to the pulse strength returned from
the fitting routine. The scaling is well described by a linear
function through the origin. } \label{rabidecay}
\end{figure*}

To study the perturbing effects of the inhomogeneous atom-probe
interaction systematically, we alternate microwave and probe pulses
and record data sets for different probe powers. In Fig.
\ref{rabidecay}a we show a collection of data together with fits of
the theoretical model from equation (\ref{phase shift}). In the
fitting model, we have allowed for a small number of spontaneous
scattering events, pumping atoms into the $6S_{1/2}(F=4, m_F \neq
0)$ states and homogeneous dephasing mechanisms like magnetic
background fluctuations, microwave driving field inhomogeneities or
cloud temperature effects \cite{Kuhr2005}.
The data is remarkably well described by the simple model. In
particular,  the envelope together with the revival of the
oscillations is very well reproduced. The fitting routine returns a
parameter $\propto \chi_0$, the maximum phase spread cause by the
light shift, which is expected to be directly proportional to the
photon number in the light--shifting pulses. The value is shown in
Fig. \ref{rabidecay}b as function of the applied photon number,
confirming the validity of our model within the given parameter
range.

\subsection{Ramsey spectroscopy}
A more direct measurement of the phase shift induced by the probing
can be obtained with Ramsey spectroscopy \cite{Featonby1998}.
\begin{figure*}[t!]\begin{center}
\includegraphics[width=2\columnwidth]{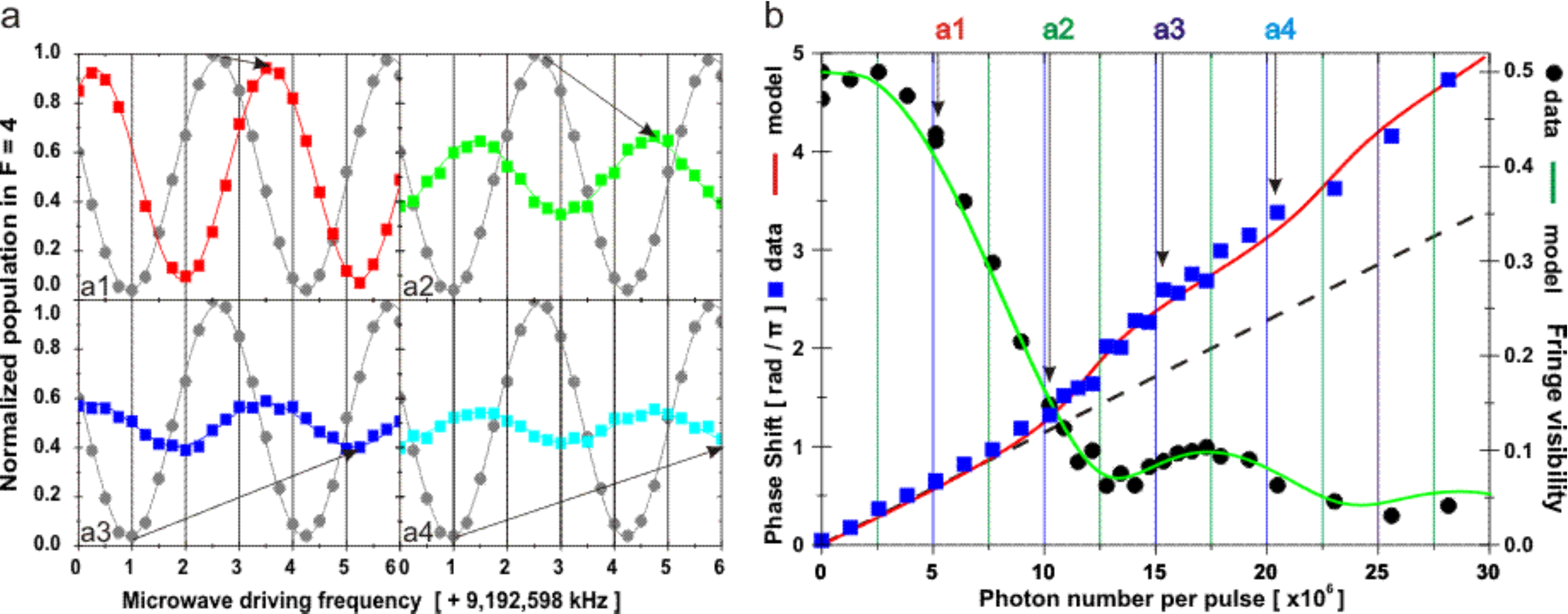}\end{center}
\caption{(a) Ramsey fringes, discrete points represent experimental
data, the solid lines are cosine fits to the data. The data
indicated with {\color{Gray} $\bullet$} represents the reference
trace where no light shifting pulse has been applied to the
superposition state during the Ramsey sequence. From (a1)--(a4), the
probe pulse photon number has been doubled in each step, starting
with $2.6 \times 10^6$. (b) Normalized phase shift and fringe
amplitude of Ramsey fringes, extracted from data similar to
(a1)--(a4). The solid lines represent the theory curves and follow
the experimental data remarkably well. The dashed line in the phase
shift data corresponds to the linear dependence expected for a
homogeneous system.} \label{ramseyprinc}
\end{figure*}
Briefly, the basic principle is as follows: Beginning from an
initial state, where all atoms reside in $|3\rangle$, a
$\pi/2$--pulse brings the ensemble into a superposition state
$\frac{1}{\sqrt{2}}(|3\rangle + |4\rangle)$. The quantum state then
evolves freely, in our case for a time of $300\,\mu$s. The
population measured in $|4\rangle$ after a second $\pi/2$--pulse
depends on the relative phase $\phi$ between the two atomic states
$\frac{1}{\sqrt{2}}(|3\rangle + e^{i \phi} |4\rangle)$ {acquired}
during the free evolution. For $\phi \mbox{ mod }2 \pi = 0$ we end
up at $|4\rangle$, $\phi \mbox{ mod } 2 \pi = \pi$ yields
$|3\rangle$ and $\phi \mbox{ mod }2 \pi = (\pi/2,3\pi/2)$ yields
$\frac{1}{\sqrt{2}}(|3\rangle \pm |4\rangle)$. As discussed in
section \ref{sec:2:1},  such a phase shift can be induced by
shifting the transition out of resonance, e.g. by applying a probe
pulse, or by detuning the driving field from resonance. The latter
yields well--known Ramsey fringes \cite{Vanier1989}, shown as
reference in Fig. \ref{ramseyprinc}a.  Again, the population
observed in $|4\rangle$ is normalized to the number of atoms. When
we, in addition, apply light shifting pulses (simply by using the
probe beam for this purpose) while the atomic state evolves freely,
the differential light shift adds a phase shift distribution
proportional to the number of photons interacting with the atoms.
Accordingly, the Ramsey fringes will be shifted in frequency space,
which can be clearly seen in graphs (a1)--(a4) of Fig.
\ref{ramseyprinc}a. By normalizing the frequency shift to the period
of the Ramsey fringes, we can directly extract the mean phase shift
angle caused by the probe. In a homogeneous system as studied by
Featonby {\it et al.} \cite{Featonby1998}, the Ramsey fringe
position  shifts proportionally to the photon number of the probe
pulse. In the inhomogeneous situation we are considering, the
spatial profile of the light pulse will create a phase shift
distribution along the equator as discussed in section
\ref{sec:2:1}. We can therefore  no longer expect the shift to be
exactly  proportional to the probe pulse strength, since states
gaining the same phase angle $\phi =( \chi t  \ \mbox{ mod } \ 2\pi)
$ are equivalent in a Ramsey experiment. The phase distribution of
the ensemble also acts to wash out the Ramsey fringe visibility,
since the externally introduced distribution is basically a standard
dephasing
mechanism.\\
In Fig. \ref{ramseyprinc}b the normalized phase shift and amplitude
of the fringes, extracted from the Ramsey spectroscopy measurements
are shown. One can see a clear deviation from a linear scaling when
the accumulated phase shift exceeds $2\pi$. The Ramsey fringe
amplitude also shows the expected revival when the phase
distribution starts to overlap above $2 \pi$ and Bloch vector
components with the same phase modulo  $2 \pi$  interfere
constructively. The graph also contains the theoretical predictions
from the model given above and a good correspondence is observed.

\begin{figure*}[!]\begin{center}
\includegraphics[width=2\columnwidth]{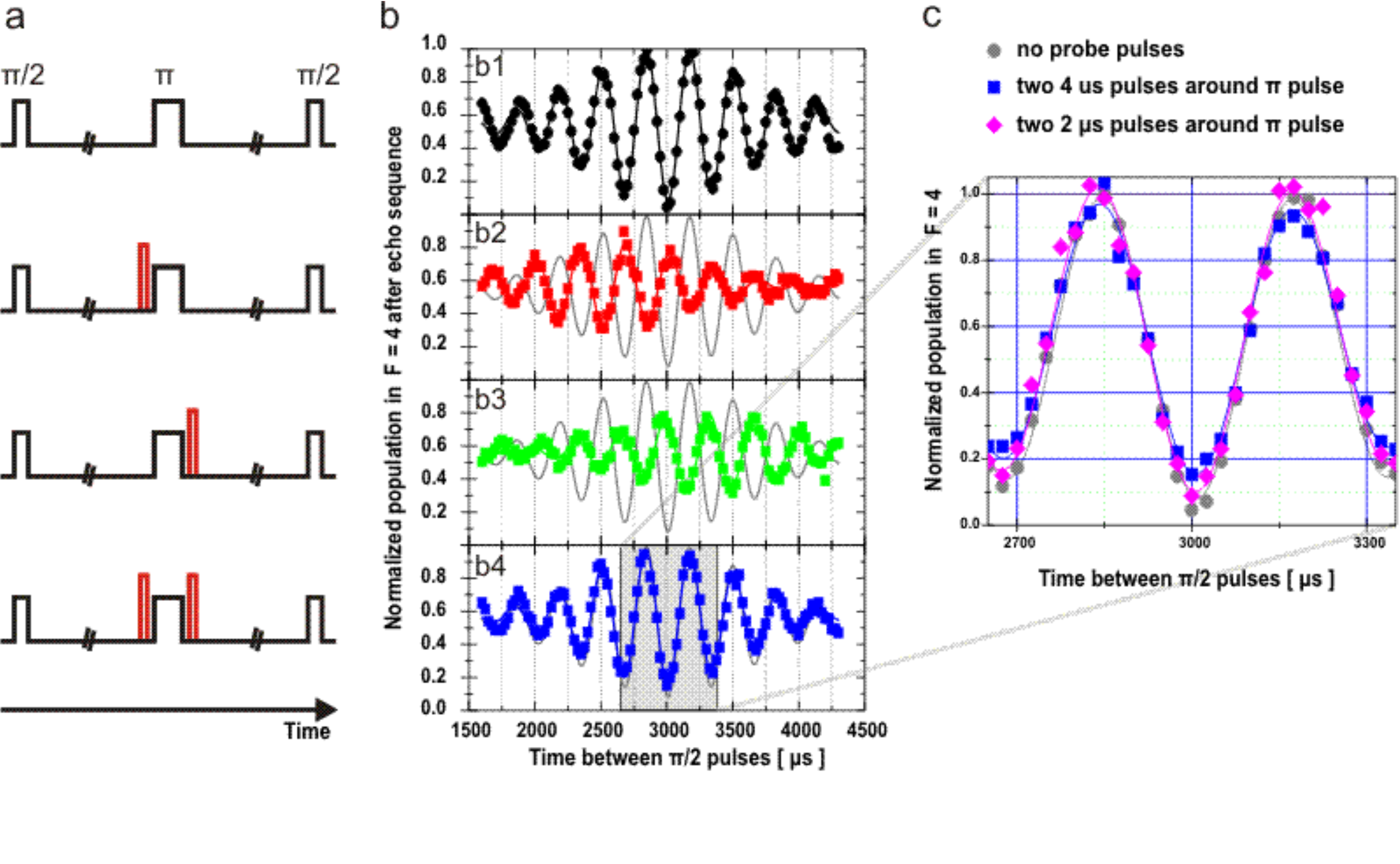}
\caption{(a) Pulse sequence for spin echo measurements. The black
lines correspond to  the microwave pulses and the red lines indicate
the placement of the probe pulses. (b) Spin echo signal observed
according to the corresponding pulse sequences to the left. The fit
to the plain spin echo trace (b1) is kept for reference in graphs
(b2)--b(4). Additional probe pulses, duration $4\,\mu$s, containing
$\sim 10^6$ photons,  during the echo sequence shift the echo signal
in time space (b2)--(b3), corresponding to the mean phase shift
imprinted onto the ensemble. Distributing the probe pulse around the
re--phasing microwave pulse, amplitude and phase of the echo signal
are regained. The solid lines represent fits to the data, assuming a
Gaussian envelope of the echo signal. (c) A zoom into the central
part of the echo signal shows that the ensemble can be fully
re--phased when the probe pulses are distributed symmetrically
around the microwave echo pulse. The slight reduction of the fringe
contrast when doubling the photon number is due to photon
scattering} \label{echo}
\end{center}\end{figure*}
\section{Sample re--phasing with spin echo}
The inhomogeneous phase spread of the ensemble after a measurement
poses a serious problem for spectroscopy, squeezing, and quantum
information applications and challenges the non--destructive nature
of the measurement. Obviously, for probe pulses with large photon
numbers, the atomic state evolution  is dominated by the effect of
the probing. Using such a strongly perturbing probe beam, e.g. to
predict the quantum state of the ensemble,  creates a state whose
phase is distributed around the equator of the Bloch sphere. When
interrogating an ensemble with $N$ atoms, one can still gain
information about the z--projection of the state to better than
$1/\sqrt{N}$ --- the standard quantum limit --- and thus achieve
spin squeezing, but due to the large phase distribution, the
remaining state will be of little use for spectroscopic
applications. It is, however, possible to re-phase the ensemble
after a dispersive measurement by applying spin echo techniques
\cite{Andersen2003}. To achieve re-phasing, we invert the time
evolution of the spreading by adding a microwave $\pi$--pulse
between the two Ramsey $\pi/2$--pulses. To again study the effect of
the differential light shift distribution when probing the sample,
light pulses are symmetrically distributed around this refocusing
$\pi$--pulse. The pulse sequence for these echo measurements is
illustrated in Fig. \ref{echo}a. After the pulse sequence, we
measure the population in $|4\rangle$ with further probe pulses and
normalize to the total number of atoms. \\
The plain spin echo sequence, taking care of dephasing e.g. caused
by the trapping laser, shows a close to perfect refocussing of the
sample at the expected time as shown in Fig. \ref{echo}b,(b1). When
applying a single probe pulse before or after the echo pulse, the
echo fringe is shifted in time according to the induced mean phase
shift by the probe pulses, figure \ref{echo}b,(b2)--(b3). As with
the Ramsey fringes in Fig. \ref{ramseyprinc}a, the  inhomogeneity of
the light shift reduces the echo fringe visibility drastically. If
we, however, apply light pulses symmetrically around the spin echo
pulse, Fig. \ref{echo}b,(b4) shows that we regain the unshifted echo
fringe almost perfectly. In graph \ref{echo}c we zoom in on the
central Ramsey fringe to back up this claim. The graph also confirms
that the measurements are indeed not limited by off--resonant photon
scattering. When the ensemble is in a  superposition state, both
inelastic Raman and elastic Rayleigh scattering would lead to
complete decoherence of the excited atoms and reduce the fringe
contrast. However, as can be seen in the graph \ref{echo}c, the
fringe contrast is, not reduced appreciably. The influence of photon
scattering is slightly visible   when comparing the fringe
amplitudes of echo signals with different probe pulse powers. From
the magnitude of the change, however, we conclude that they are of
minor consideration here. In addition, the data confirms that other
dephasing mechanisms e.g. due to the non--zero temperature of the
cloud or magnetic background fluctuations  are also of minor
importance.

\section{Discussion}
We use a dispersive phase shift measurement to non-- destructively
probe the populations of the clock states in a cold Cs ensemble when
subjected to near resonant microwave pulses. The strong dependence
of the Rabi oscillation  envelope on the probe pulse photon numbers
--- causing a differential light shift between the clock states  and
thus adding a relative phase between the clock states --- can only
be explained by the inhomogeneity of the light--atom interaction,
which is intrinsic to the experimental setup. With the introduced
theoretical model, the experimental data can be convincingly
explained.  To quantify the dephasing of the atomic ensemble induced
by the probing, Ramsey spectroscopy in the frequency domain is used,
and both the scaling of the phase shift of the Ramsey fringes and
the scaling of their amplitude with applied photon number in the
probe pulses can be explained by the same model. Finally, the
application of spin echo techniques allow us to re--phase the atomic
sample.\\
We note that for the range of probe intensities applied in the
experiments presented in the current paper, the effect of
spontaneous photon scattering from the probe pulses can be
neglected. Furthermore, other decoherence effects e.g. due to finite
temperature of the ensemble or magnetic field fluctuations can be
disregarded on the given timescales. However, to achieve
considerable spin squeezing in the presented experimental
configuration, considerably higher probe powers implying about
20\,\% spontaneous scattering probability are necessary
\cite{Hammerer2004}. When other decoherence effects are under
control, spin echo techniques can be used to calibrate these effects
as well \cite{Oblak2008}.

The differential light shift between the clock states due to the
probe light is, of course, caused by the choice of the probe
detuning. By choosing  a ``magic'' frequency for the probe, where
both clock levels are shifted by the same amount,  the induced
dephasing of the two levels can be minimized and in the ideal case,
canceled \cite{Chaudhury2006}. Since the presented single frequency
probing scheme is only sensitive to the scalar polarizability of the
atoms, the magic frequencies coincide with the probe detunings where
the interferometer phase shift is insensitive to the population
number difference. Single frequency measurements are thus not suited
to eliminate the perturbation of the atomic levels. The problem can
be circumvented by adding a second probe frequency or invoking the
tensor polarizability in off--resonant polarization measurements
\cite{Saffman2007}.\\

This work was funded by the Danish National Research Foundation, as
well as the EU grants QAP and COVAQUIAL. N.K. acknowledges the
support of the Danish National Research Council through a Steno
Fellowship. We would like to thank J\"org Helge M\"uller for
stimulating discussions.


\begin{thebibliography}{34}

\bibitem{Sadler2006}
L.E. Sadler, J.M. Higbie, S.R. Leslie, M.~Vengalattore,
D.~Stamper-Kurn, Nature
  \textbf{443}, 312 (2006)

\bibitem{Petrov2007}
P.G. Petrov, D.~Oblak, C.L. Garrido~Alzar, N.~Kj{\ae}rgaard, E.S.
Polzik, Phys.
  Rev. A \textbf{75}, 033803 (2007)

\bibitem{Chaudhury2006}
S.~Chaudhury, G.A. Smith, K.~Schulz, P.S. Jessen, Phys. Rev. Lett.
\textbf{96},
  043001 (2006)

\bibitem{Usami2007}
K.~Usami, M.~Kozuma, Phys. Rev. Lett. \textbf{99}, 140404 (2007)

\bibitem{Windpassinger2007}
P.J. Windpassinger, D.~Oblak, P.G. Petrov, M.~Kubasik, M.~Saffman,
C.L.
  Garrido~Alzar, J.~Appel, J.~M\"uller, N.~Kj{\ae}rgaard, E.S. Polzik, Phys.
  Rev. Lett. \textbf{100}, at press ({\it Preprint} arXiv:0801.4126) (2008)

\bibitem{Kuzmich1998}
A.~Kuzmich, N.P. Bigelow, L.~Mandel, EuroPhys. Lett. \textbf{42},
481 (1998)

\bibitem{Julsgaard2001}
B.~Julsgaard, A.~Kozhekin, E.S. Polzik, Nature \textbf{413}, 400
(2001)

\bibitem{Geremia2004}
J.~Geremia, J.~Stockton, H.~Mabuchi, Science \textbf{304}, 270
(2004)

\bibitem{Sherson2006}
J.~Sherson, B.~Julsgaard, E.~Polzik, Adv. At. Mol. Opt. Phys.
\textbf{54}
  (2006)

\bibitem{Hammerer2004}
K.~Hammerer, K.~M\o{}lmer, E.S. Polzik, J.I. Cirac, Phys. Rev. A
\textbf{70},
  044304 (2004)

\bibitem{Julsgaard2004}
B.~Julsgaard, J.~Sherson, J.I. Cirac, J.~Fiurasek, E.S. Polzik,
Nature
  \textbf{432}, 482 (2004)

\bibitem{ShersonNat2006}
J.F. Sherson, H.~Krauter, R.K. Olsson, B.~Julsgaard, K.~Hammerer,
I.~Cirac,
  E.S. Polzik, Nature \textbf{443}, 557 (2006)

\bibitem{Santarelli1999}
G.~Santarelli, P.~Laurent, P.~Lemonde, A.~Clairon, A.G. Mann,
S.~Chang, A.N.
  Luiten, C.~Salomon, Phys. Rev. Lett. \textbf{82}, 4619 (1999)

\bibitem{Wineland1992}
D.J. Wineland, J.J. Bollinger, W.M. Itano, F.L. Moore, D.J. Heinzen,
Phys. Rev.
  A \textbf{46}, R6797 (1992)

\bibitem{Wineland1994}
D.J. Wineland, J.J. Bollinger, W.M. Itano, D.J. Heinzen, Phys. Rev.
A
  \textbf{50}, 67 (1994)

\bibitem{Oblak2005}
D.~Oblak, P.G. Petrov, C.L. Garrido~Alzar, W.~Tittel, A.K.
Vershovski, J.K.
  Mikkelsen, J.L. S{\o}rensen, E.S. Polzik, Phys. Rev. A \textbf{71}, 043807
  (2005)

\bibitem{Meiser2007}
D.~Meiser, J.~Ye, M.J. Holland, {\it Preprint} arXiv:0707.3834

\bibitem{Takamoto2005}
M.~Takamoto, F.L. Hong, R.~Higashi, H.~Katori, Nature \textbf{435},
321 (2005)

\bibitem{Targat2006}
R.~Le~Targat, X.~Baillard, M.~Fouch\'{e}, A.~Brusch, O.~Tcherbakoff,
G.D.
  Rovera, P.~Lemonde, Phys. Rev. Lett. \textbf{97}, 130801 (2006)

\bibitem{Ludlow2006}
A.D. Ludlow, M.M. Boyd, T.~Zelevinsky, S.~Foreman, S.~M.and~Blatt,
M.~Notcutt,
  T.~Ido, J.~Ye, Phys. Rev. Lett. \textbf{96}, 033003 (2006)

\bibitem{Featonby1998}
P.D. Featonby, C.L. Webb, G.S. Summy, C.J. Foot, K.~Burnett, J.
Phys. B
  \textbf{31}, 375 (1998)

\bibitem{Vanier1989}
J.~Vanier, C.~Audoin, \emph{The Quantum Physics of Atomic Frequency
Standards}
  (Adam Hilger, 1989)

\bibitem{Grimm2000}
R.~Grimm, M.~Weidem\"uller, Y.B. Ovchinnikov, Adv. At. Mol. Opt.
Phys.
  \textbf{42}, 95 (2000)

\bibitem{AVILA1987}
G.~Avila, V.~Giordano, V.~Candelier, E.~Declercq, G.~Theobald,
P.~Cerez, Phys.
  Rev. A \textbf{36}, 3719 (1987)

\bibitem{Tremblay1990}
P.~Tremblay, C.~Jacques, Phys. Rev. A \textbf{41}, 4989 (1990)

\bibitem{Tannoudji1977}
C.~Cohen-Tannoudji, B.~Diu, F.~Lalo\"e, \emph{Quantum Mechanics}
(Wiley, New
  York, 1977)

\bibitem{Loudon1973}
R.~Loudon, \emph{The Quantum Theory of Light} (Oxford University
Press, 1973)

\bibitem{Ozeri2005}
R.~Ozeri, C.~Langer, J.D. Jost, B.~DeMarco, A.~Ben-Kish, B.R.
Blakestad,
  J.~Britton, J.~Chiaverini, W.M. Itano, D.B. Hume et~al., Phys. Rev. Lett.
  \textbf{95}, 030403 (2005)

\bibitem{Metcalf1999}
H.~Metcalf, P.~van~der Straten, \emph{Laser Cooling and Trapping}
(Springer,
  Berlin, 1999)

\bibitem{Allen1987}
L.~Allen, J.H. Eberly, \emph{Optical Resonance and Two-Level Atoms}
(Dover, New
  York, 1987)

\bibitem{Kuhr2005}
S.~Kuhr, W.~Alt, D.~Schrader, I.~Dotsenko, Y.~Miroshnychenko,
  A.~Rauschenbeutel, D.~Meschede, Phys. Rev. A \textbf{72}, 023406 (2005)

\bibitem{Andersen2003}
M.F. Andersen, A.~Kaplan, N.~Davidson, Phys. Rev. Lett. \textbf{90},
023001
  (2003)

\bibitem{Oblak2008}
D.~Oblak, {\it et al.}, in preparation  (2008)

\bibitem{Saffman2007}
M.~Saffman, {\it et al.}, in preparation  (2007)

\end{thebibliography}

\end{document}